\documentclass[
 prapplied,
 amsmath,amssymb,
 reprint, longbibliography, superscriptaddress
]{revtex4-2}

\usepackage{graphicx}
\usepackage{dcolumn}
\usepackage{bm}
\usepackage[utf8]{inputenc}
\usepackage[T1]{fontenc}
\usepackage{mathptmx}
\usepackage{etoolbox}
\usepackage{amssymb}
\usepackage{amsmath}
\usepackage{hyperref}
\usepackage[version=3]{mhchem}
\usepackage{esint}
\usepackage{gensymb}
\usepackage{url}
\usepackage{mathptmx}
\usepackage{siunitx}

\begin{document}

\preprint{APS/123-QED}

\title[Surface Modification and Coherence in Lithium Niobate SAW Resonators]{Surface Modification and Coherence in Lithium Niobate SAW Resonators}

\author{Rachel G. Gruenke}
\email{rgruenke@stanford.edu}
\author{Oliver A. Hitchcock}
\affiliation{ 
Department of Applied Physics and Ginzton Laboratory, Stanford University, 348 Via Pueblo Mall, Stanford, California 94305, USA
}
\author{E. Alex Wollack}
\affiliation{AWS Center for Quantum Computing, Pasadena, California 91106, USA}

\author{Christopher J. Sarabalis}
\affiliation{ Flux Photonics, Inc., 580 Crespi Dr \# R, Pacifica, CA 94044, USA}

\author{Marc Jankowski}
\author{Timothy P. McKenna}
\affiliation{NTT Research Inc., Physics and Informatics Laboratories, 940 Stewart Dr., Sunnyvale, CA 94085, USA}

\author{Nathan R. Lee}
\author{Amir H. Safavi-Naeini}
\email{safavi@stanford.edu}
\affiliation{ 
Department of Applied Physics and Ginzton Laboratory, Stanford University, 348 Via Pueblo Mall, Stanford, California 94305, USA
}

\date{\today}

\begin{abstract}
Lithium niobate is a promising material for developing quantum acoustic technologies due to its strong piezoelectric effect and availability in the form of crystalline thin films of high quality. However, at radio frequencies and cryogenic temperatures, these resonators are limited by the presence of decoherence and dephasing due to two-level systems. To mitigate these losses and increase device performance, a more detailed picture of the microscopic nature of these loss channels is needed. In this study, we fabricate several lithium niobate acoustic wave resonators and apply different processing steps that modify their surfaces. These treatments include argon ion sputtering, annealing, and acid cleans. We characterize the effects of these treatments using three surface-sensitive measurements: cryogenic microwave spectroscopy measuring density and coupling of TLS to mechanics, x-ray photoelectron spectroscopy and atomic force microscopy. We learn from these studies that, surprisingly, increases of TLS density may accompany apparent improvements in the surface quality as probed by the latter two approaches. Our work outlines the importance that surfaces and fabrication techniques play in altering acoustic resonator coherence, and suggests gaps in our understanding as well as approaches to address them.  
\end{abstract}

\maketitle

\section{Introduction}
Mechanical resonators oscillating at radio frequencies (RF) hold promise as components capable of performing important memory, processing, and transduction functions in emerging quantum systems \cite{o2010quantum, chu2017quantum, chu2018creation, satzinger2018quantum, sletten2019resolving, hann2019hardware, wollack2021loss, wollack2022quantum, qiao2023developing, bild2023schrodinger}. However, like their superconducting counterparts, these RF acoustic systems are subject to two-level system (TLS) loss at low temperatures \cite{wollack2021loss, andersson2021acoustic, cleland2023studying, luschmann2023surface}, which limits their lifetime and utility. The standard tunneling model for two-level systems describes the temperature-dependent loss, frequency shift, and saturation behaviors \cite{gao2008experimental, behunin2016dimensional} accurately, albeit without a precise microscopic description. It is therefore unclear how different materials and fabrication approaches affect the two-level systems in a material, and their effect on the cryogenic mechanical properties. 

This study systematically explores the effect of several typical acoustic resonator fabrication steps on the density of TLS in Lithium Niobate (LN). We use a series of fabricated surface acoustic wave resonators (SAWs) on Lithium Niobate to study surface TLS sources in RF mechanical resonators. We choose LN due to its growing application as a material for implementing hybrid quantum systems, and SAWs as opposed to other types of acoustic waves because the driven mechanical motion overlaps strongly with the surface of the substrate, making them a suitable probe for surface-lying TLS sources. Moreover, they are easy to fabricate with very few processing steps, as compared to other acoustic resonators. A single metalization step on bulk LN is enough to realize a resonator, significantly reducing excess processing that could alter and complicate the final TLS density. We fabricate several resonators with different surface preparation steps and measure their TLS density at cryogenic temperatures. To develop a better understanding of the surface chemical composition, we use x-ray photoelectron spectroscopy (XPS). We also use atomic force microscopy (AFM) to investigate surface topography changes. Through this process, we identify surface treatments that increase TLS density, such as annealing or ion milling, and other surface treatments that do not alter the TLS density, such as piranha dipping, and correlate these changes with modifications of the surface properties.

\section{Fabrication and Surface Treatment}
\begin{figure*}[ht]
\includegraphics[width=\linewidth]{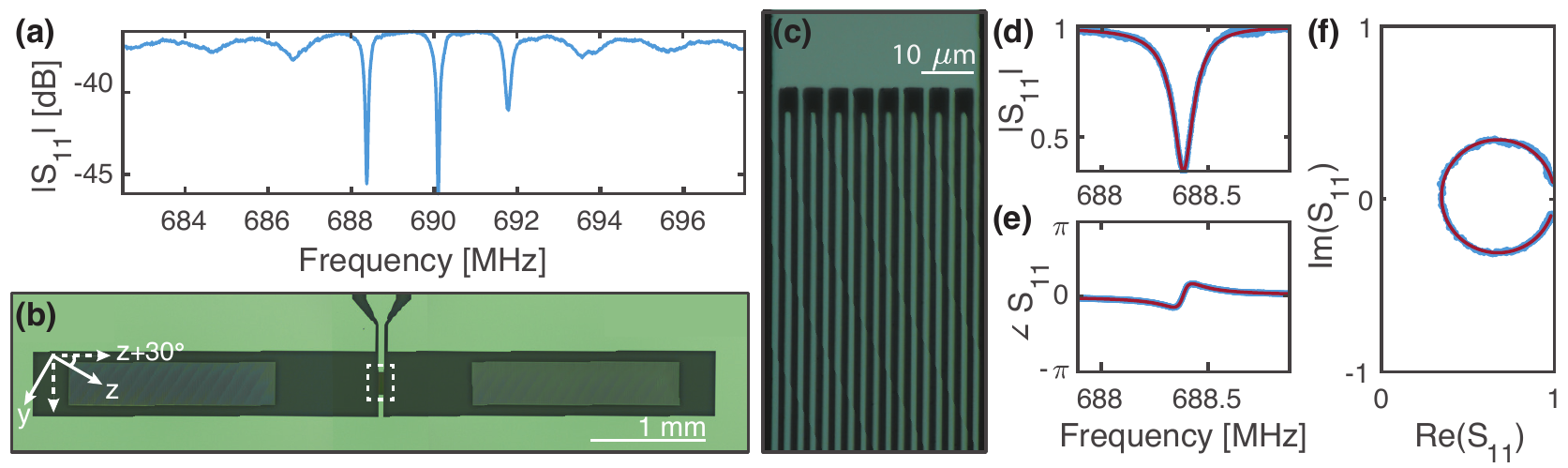}
\caption{\label{fig:fig1} \emph{Overview of the SAW devices} (a) The magnitude of $S_{11}$ across the full stopband of the device taken at 10~mK. Modes are centered at 690~MHz with a 1.7~MHz FSR. (b) - (c) Optical microscope images of the SAW features: image (b) details the full SAW device with central IDT and Bragg mirrors on either side, image (c) zooms into the IDT fingers. (d)-(f) Zoomed in reflection measurement (blue) and fit to a Lorentzian (red) of the first SAW mode centered at 688.4~MHz. Magnitude (d), phase (e) and real vs. imaginary parts (f) of $S_{11}$ are plotted for the same mode. Fit to the mode finds $Q_i = 6.8\times10^3$ and $Q_e = 1.4\times10^4$.}
\end{figure*}

To understand the changes in the TLS-induced loss, we fabricate several SAW resonators on bulk LN, introducing varying surface treatment steps between substrate preparation and metalization. We start with a 500~{\textmu}m thick bulk x-cut LN substrate from Precision Micro Optics. All but one device use congruently grown LN substrates; the other uses a 5\% Magnesium Oxide (MgO) co-doped LN substrate. We prepare substrate pieces by dicing the vendor material and sonicating each piece in first acetone and then isopropanol. 

Once the substrates pieces are individually solvent cleaned, we perform different surface treatments on each substrate. The "CLN" and "MgO" devices undergo no additional surface treatment- the former serving as our control group and the latter testing the effects of LN stoichiometry on TLS density. We anneal the "Annealed" devices at 500°C for eight hours in an ambient environment. We dip the "BOE" devices in 6:1 concentration buffered oxide etchant (BOE) for 2 minutes following a 500°C anneal. We submerge "Piranha" devices in a 3:1 sulfuric acid:hydrogen peroxide concentration solution for 20-minutes. Finally, the "GCIB" devices feature no additional surface treatment until after metalization. 

Once we complete surface treatments of each type, we define the SAW fingers, terminals and ground planes using photolithography and liftoff on some of the treated substrates of each type. We spin a bilayer of LOR5A and SPR3612 photoresists and expose our features using a Heidelberg MLA 150. We develop in MF26A and then deposit 25~nm thick aluminum in our Plassys evaporator. We finish the SAW metalization by lifting off in NMP solvent. The "GCIB" devices undergo a final three-minute 10kV 30nA ion sputter from a Gas Cluster Ion Beam to study effects of organics removal after metalization. We then package the final SAW devices for cryogenic testing in PCBs and copper enclosures. Surface treated substrates that did not undergo metalization are reserved for XPS and AFM measurements. 

The SAW devices we pattern on all substrates consist of an inter-digitated transducer (IDT) and two Bragg mirrors, defining an acoustic cavity. The IDT consists of 17 alternating 200~{\textmu}m long and 1~{\textmu}m wide fingers, with a 40\% duty cycle. The Bragg mirrors have 700 400-{\textmu}m-long and 1.5-{\textmu}m-wide fingers, with a 60\% duty cycle. These parameters are chosen to optimize our stop band and free spectral range (FSR), yielding three well separated SAW modes for TLS characterization. To minimize the effect of diffraction loss caused by beam-steering in anisotropic LN and improve the internal quality factors \cite{emser2022minimally}, the IDT fingers are patterned such that the SAW drive direction is parallel to the crystal Z+30\degree of LN, as seen in Fig.~\ref{fig:fig1}. We discuss more details on this aspect of the design in Appendix \ref{walkoff}.

\begin{table}[ht]
\caption{\emph{Surface Treatment Steps for Each Device}}
\centering
\begin{ruledtabular}
\begin{tabular}{l l}
\textbf{Device} & \textbf{Treatment Steps} \\
CLN & None (Control) + metalization \\
Annealed & 500°C anneal, 8h + metalization \\
BOE & 500°C anneal, 8h + 2 min BOE dip + metalization \\
MgO & 5\% MgO co-doped substrate + metalization \\
Piranha & 20 min Piranha + metalization\\
GCIB & Metalization + 3 min 10kV GCIB \\
\end{tabular}
\end{ruledtabular}
\end{table}

\section{TLS Measurement}

The product \(F \delta_{\text{TLS}}^0\) is a key parameter in quantifying the impact of TLS. The larger its value, the greater the participation of TLS loss channels, and hence, the higher the system's loss due to TLS. Here, \(F\) denotes the filling fraction of TLS in the mode volume, which indicates the proportion of the volume occupied by the TLS. \(\delta_0\) is the average TLS loss tangent, a measure of energy dissipation in the system due to the TLS. Our goal is to determine  \(F \delta_{\text{TLS}}^0\). Though it is possible to do this by looking at changes in the internal quality factor of the SAW modes, we choose to measure the temperature-dependent modal frequency shift:
\begin{align}\label{eq:one}
    \!\!\!\!\!\frac{\Delta\omega_r}{\omega_r} = \frac{F\delta_{\text{TLS}}^0}{\pi}\! \left[ \text{Re}\!\left\{\!\Psi\!\left(\frac{1}{2}\!+\!\frac{\hbar\omega_r}{2\pi i k_B T}\right)\!\right\} \!-\!\ln\frac{\hbar\omega_{r}}{2\pi k_B T}\right].\!\!\!\!\!\!
\end{align}
Here, $\Delta \omega_r$ is the frequency shift of the mechanical oscillator from its nominal frequency $ \omega_r$ at 200~mK, and $T$ is the temperature. To perform these measurements, we first package and mount the devices on the base plate of a Bluefors LD400 dilution refrigerator, where they may be cooled to a base temperature below 10~mK. The setup includes a total of $-66$~dB of RF attenuation in the fridge distributed between the different temperature stages to remove most of the thermal photons in the band of interest at the device input. To obtain the reflection signal, we place two cryogenic circulators directly before the device, send in a microwave signal and measure the reflected signal on the reflection port. We fit the resulting spectra to obtain the resonance frequency as well as the intrinsic and extrinsic quality factors for all modes in our SAW stop band. 

At $10$~mK, our SAW design yields a nominal device with a 5~MHz stop band centered at 690~MHz. The mirrors are separated by 1.75~mm giving the SAW modes an FSR of 1.7~MHz; thus, we can resolve around three SAW modes per device. At high power, we find that the SAW modes' internal and external quality factors across different surface-treated substrates were within the range of $Q_i = 2\times10^3 - 12\times10^3$ and $Q_e = 1\times10^4 - 3\times10^4$. All modes are undercoupled. Although the internal quality factors exhibit loss due to TLS, as seen in temperature sweep measurements detailed below, the losses are likely still dominated by metal or diffraction loss. 

To extract \(F \delta_{\text{TLS}}^0\), we sweep the SAW temperature from 10~mK to 200~mK using a resistive heater placed at the same temperature stage as the device; we apply the heat constantly and allow the devices to equilibrate before measurement. At each temperature, we measure the resonances and fit the resonance frequency $f_0$. As the SAW temperature decreases from a reference temperature at 200~mK to the base temperature, the resonance frequency of each mode decreases, a specific signature of low-temperature TLS loss \cite{gao2008experimental, wollack2021loss}. For each device, we fit the temperature-dependent resonance frequency shifts to the model in equation \ref{eq:one}. 

\begin{figure}[t!]
\includegraphics[width=\linewidth]{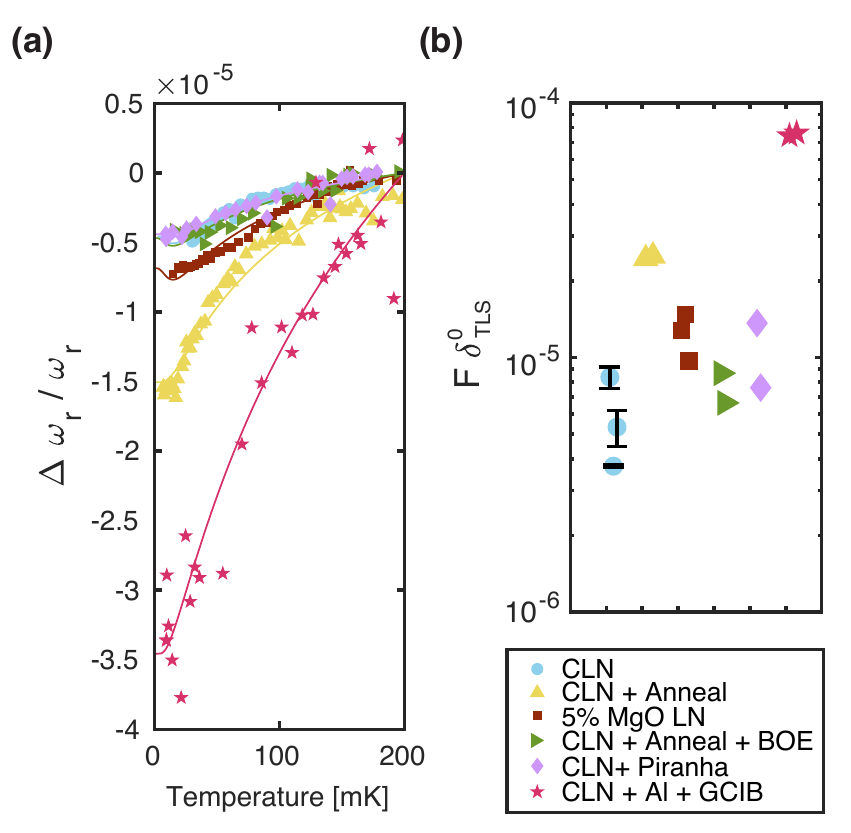}
\caption{\label{fig:fig2} \emph{Fit SAW redshift measured from T = 200~mK to 10~mK} (a) The decreased resonant frequency of different SAW modes, and each data sets' fit to equation \ref{eq:one}. The modes on SAW devices that were argon ion milled with GCIB prior to cooldown had the largest redshift. The devices that had been annealed at 500C for 8hr at atmospheric pressure had the next largest redshift, followed by the 5\% MgO co-doped device. All other surface treatment steps yielded similar low temperature redshift. (b) Fitted $F \delta^0_\text{TLS}$ products. Error bars shown on the CLN device are defined by measuring a single device multiple times per cooldown and recording changes in the redshift fit.}
\end{figure}

We compare the fit scaling factor \(F \delta_{\text{TLS}}^0\) for modes on each surface treated SAW~\cite{pappas2011two}, as shown in  Fig.~\ref{fig:fig2}. The addition of GCIB sputtering yields the largest $F \delta^0_\text{TLS}$  at $7.6\times10^{-5}$, more than 14 times larger than the control CLN $F \delta^0_\text{TLS}$. Annealing in atmosphere without a subsequent BOE dip increases the $F \delta^0_\text{TLS}$ significantly as compared to CLN TLS loss; these modes have a TLS loss participation fraction $\approx 5\times$ larger than the CLN modes. 5\% MgO co-doped devices have a slightly elevated TLS loss participation, $\approx 2\times$ times larger than CLN. All other types of surface preparation devices have average TLS $F \delta^0_\text{TLS}$ similar to CLN within the error bar bounds, set by standard deviation between consecutive measurements of the same modes. 

We find that the $Q_i$'s are sufficiently large to resolve the TLS loss channels in temperature sweep measurements when undercoupled. The smallest $Q_i$ we can use to detect TLS losses through temperature sweeping is set by the signal to noise ratio of the SAW mode and fit error of the resonance frequency -- the resonance frequency fit error must be smaller than the total frequency redshift. We also study the TLS saturation behavior by sweeping the RF power sent to the SAW resonators. However, for TLS saturation to be easily resolved through swept RF power, $Q_\text{TLS}$ must dominate the losses of the acoustic mode. We calculate our devices' quality factors from only TLS losses at low RF power from our fitted $F \delta_0$ using the following equation: 
\begin{equation}
    Q_\text{TLS} = \frac{1}{F \delta^0_\text{TLS} \tanh{\hbar \omega / 2 k_B T}},
\end{equation}
where $\omega$ is the resonance frequency at base temperature and T is base temperature at 10~mK. We see that while the fitted total internal quality factor ranges between $Q_i = 2\times10^3 - 12\times10^3$, $Q_\text{TLS} = 1\times10^4 - 3\times10^5$, an order of magnitude larger that the total $Q_i$. Therefore, other losses such as diffraction or metal losses dominate the $Q_i$ of our SAW modes, making power sweep TLS saturation effects less effective for extracting TLS density. Further details on power sweep measurements can be found in Appendix \ref{powersweeps}.

\begin{figure*}[t!]
\includegraphics[width=\linewidth]{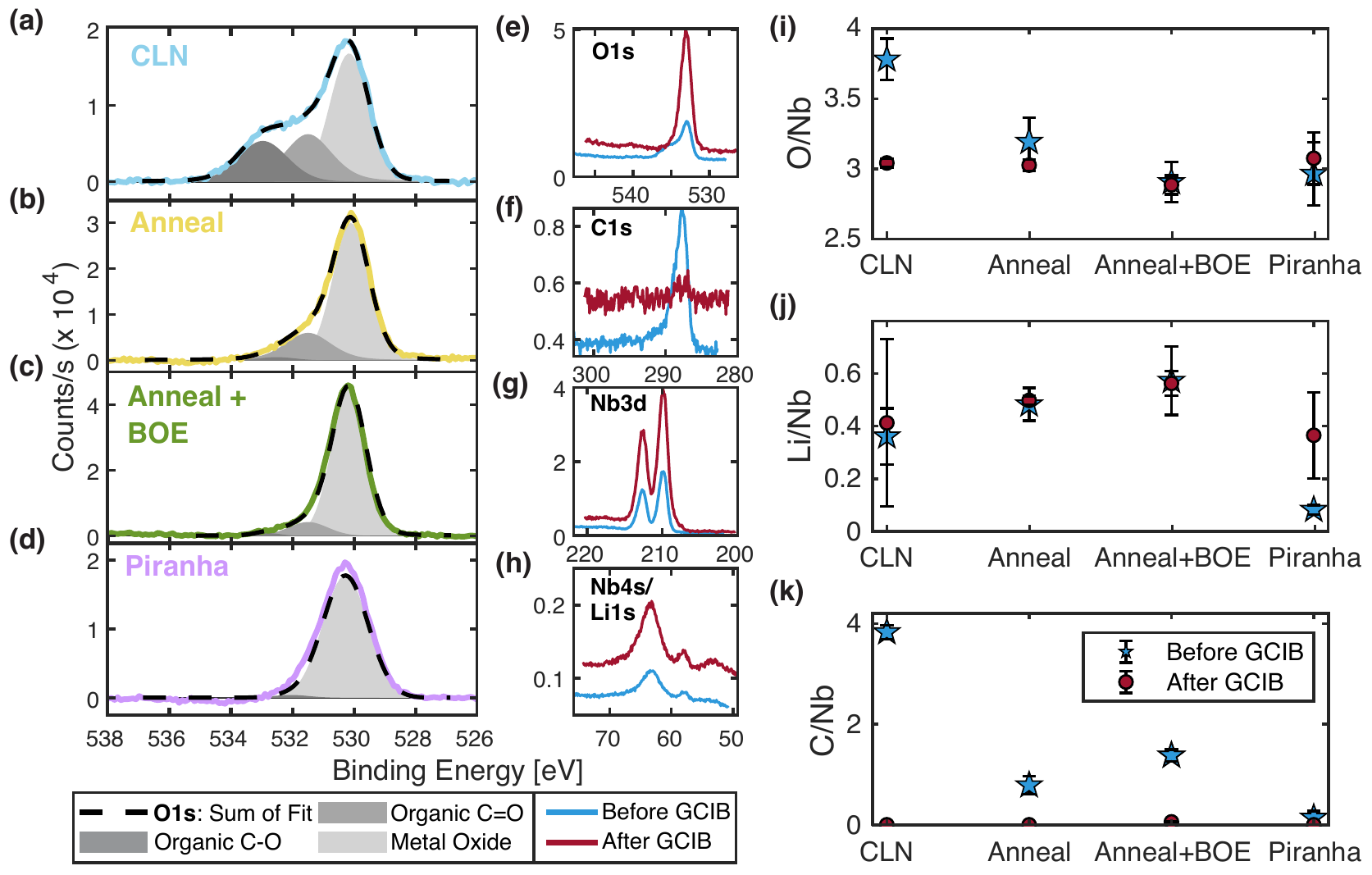}
\caption{\label{fig:fig3} \emph{XPS results comparing various LN surface treatments} (a) - (d) The O$1s$ peak of CLN, annealed LN, annealed and BOE cleaned LN, and piranha submerged LN. Each peak is compensated from charging by shifting the measured Nb peak to LN Nb$3d$ peak value of 207.3~eV; the background counts have also been subtracted from each peak. Fits to three Gaussian-Lorentzian bands for three kinds of oxygen bonding are plotted in grey: metal oxide centered at 530~eV in light grey, organic \ce{C\bond{=}O} bonding at 531.5~eV in medium grey, and organic \ce{C\bond{-}O} bonding at 533~eV in dark grey. Peaks with less of a carbon presence have negligible organic carbon peak bands, such as the case with piranha (d). (e)-(h) The four primary elemental peaks (O$1s$, C$1s$, Nb$3d$, and Li$1s$) of LN before (blue) and after (red) ion sputtering the surface with a Gas Cluster Ion Beam. Note that the smaller, narrower Li$1s$ peak at 54.8eV overlaps with Nb$4s$, which resides at 60.2eV. Removal of surface lying organics can be seen with the near annihilation of the C$1s$ peak, as well as narrowing of the O$1s$ peak. (i)-(k) The relative atomic percentage ratios of each measured surface treated sample are calculated by integrating the counts in each peak. Percentages are averaged across a 10mm chip on several samples of each surface preparation type. Black lines represent the error bars calculated from the standard deviation between measurements of the same surface treatment.}
\end{figure*}

\section{X-ray Photoelectron Spectroscopy}

X-ray photoelectron spectroscopy (XPS) is a powerful analytical technique used for understanding the chemical composition of surfaces. Exploiting the photoelectric effect, an XPS measurement bombards the sample with x-rays of a known energy and identifies chemical composition of the sample from the calculated kinetic energy of emitted photoelectrons. In the context of this study, XPS provides valuable insights into the atomic and molecular constituents of the substrate surfaces, their bonding states, and how these properties change with different surface treatments. Our objective is to discern the correlation between TLS density and  modifications in atomic composition of the surfaces brought about by different surface treatments. We map out the atomic composition of the surfaces by performing XPS analysis (PHI Versaprobe III XPS) of the surface of each treated substrate on ``twin'' chips that are not metalized.  Collected photoelectrons only escape from the top 5~nm of the substrate, so this approach probes the surface chemistry changes due to surface treatment. Each XPS measurement averages a 200~$\mu\text{m}$ diameter spot; we report average stoichiometry of each sample type by measuring multiple points across a 10mm chip.

We also use an in-situ gas cluster ion beam (GCIB) for surface cleaning and depth profiling of samples; GCIB is advantageous for depth profiling in XPS as compared to a standard argon ion sputtering as the beam is less damaging to the sample bulk. We measure each surface-treated sample before and after an in-situ 3~min 10~kV GCIB sputter. This allows us to analyze how the surface treatments change the top 5~nm of material and verify there are no additional changes to the bulk LN stoichiometry. 

We analyze the XPS spectra to obtain the average atomic percentage by summing all integrated peak counts. Each measurement collects the four strongest spectral lines for LN with surface lying contaminants: carbon C$1s$, oxygen O$1s$, niobium Nb$3d$, and lithium Li$1s$. To compensate for peak shifts due to sample charging, we shift each peak with the reference Nb$3d$5/2 peak at 207.3~eV for lithium niobate \cite{kaufherr1996x, nist2000nist}, as it is the most prominent niobium peak of LN and it does not shift in the presence of surface organics; we choose to not shift peaks with the adventitious carbon peak reference, as this peak is removed after GCIB sputtering. We fit the total counts using a Shirley background subtraction and peak fitting (Multipak software). 

The average atomic percentage for a single element $x$ is given by 
\begin{equation}
C_{x} = \frac{F_x^{-1}\displaystyle\int I_x \, dBE}{ \displaystyle\sum_i F_{i}^{-1} \displaystyle\int I_{i} \, dBE \, }, 
\end{equation}
where $\int I_i dBE$ is the integrated counts of the measured elemental line after Shirley background subtraction over the binding energy, and $F_i$ is the tool specific atomic sensitivity factor, which is directly proportional to the photoabsorption cross-section \cite{chastain1992handbook}. To compare various spectra with the quantity best representative of stoichiometry, we report the ratios of carbon, lithium and oxygen atomic percentage to the niobium percentage. A summary of XPS atomic percentage ratio results before and after GCIB sputtering for CLN, annealed, BOE, and piranha samples can be seen in Fig.~\ref{fig:fig3} (i)-(k). Black error bars represent the standard deviation of atomic percentage ratios after multiple XPS measurements of each sample type. The MgO and GCIB samples are not included in this comparison as these samples have extra atomic species (magnesium in MgO samples from co-doping, and aluminum in GCIB because this surface treatment occurred after metalization), meaning that the atomic percentage ratios are not easily compared to CLN substrates.

Before performing GCIB, the CLN, Anneal, and BOE samples all have excess carbon, compared to the post-GCIB samples where carbon is undetectable. The piranha treatment nearly removes all carbon contaminants; C/Nb = 0.14 for piranha, as compared to C/Nb = 3.82 for CLN and C/Nb = 0.78 for anneal. CLN and annealed samples also have excess oxygen that is removed after GCIB. CLN has the largest relative ratios of oxygen and carbon. These relative ratios decrease from CLN levels after anneal and after anneal + BOE treatments. The measured lithium content does not vary much for different surface treatment.

The atomic percentages after the in-situ GCIB shows the efficacy of GCIB in standardizing the stoichiometry across different samples after the removal of adsorbed surface layers. We find that the samples after GCIB all have very similar stoichiometry: oxygen at $67 \pm 2 \%$, niobium at $22.3 \pm 0.2 \%$, lithium at $10 \pm 2 \%$, and carbon being undetectable. Note that there is a large systematic offset from the calculated lithium percentage and expected CLN stoichiometry; the average Li/Nb is $0.46 \pm 0.09$, when CLN should have a ratio of Li/Nb lightly less than 1 (0.95) \cite{carruthers1971nonstoichiometry, rauber1978current}. We believe this to be due to systematic error caused by the the small photoabsorption cross-section of Li$1s$ and subsequently the low intensity of the Li$1s$ spectral peak. However, variations in the Li/Nb and O/Nb ratios from sample to sample after GCIB are much smaller than before GCIB. This suggests that differences in the measured TLS densities may be due to variations in adsorbed oxygen and carbon on the top few nanometers of CLN, annealed, and BOE dipped samples, rather than by significant changes in the bulk of the crystal. 

We also use XPS spectra to study chemical bonding type via peak line shape. The bonding type of surface lying oxygen can be determined from the O$1s$ spectral line shape. When there is excess organic contaminant on the surface from residual resists, higher energy carbon bonding bands at 533~eV and 531.5~eV take up a larger percentage of the total O$1s$ counts \cite{naumkin2012nist, zhang2008surface, yang2009chemical}. On the other hand, the primary oxygen bonding in lithium niobate is an $\text{Nb}_2\text{O}_5$ metal oxide centered at 530.5~eV. The changes in oxygen bonding bands can be seen in Figure~\ref{fig:fig3} (a) - (d). CLN control samples which have no additional cleaning after removal of dicing resist via solvents have the largest presence of \ce{C\bond{-}O} and \ce{C\bond{=}O} oxygen bonding. Piranha dipped LN samples target and remove organics, such that the organic carbon bonding bands are absent in this sample. Annealed and annealed plus BOE cleaned samples have a larger prominence of metal oxide bonding than CLN, but still show some organic oxygen bonding in the \ce{C\bond{=}O} shoulder. 

Similarly, figures~\ref{fig:fig3} (e)-(h) show how the all the spectral lines change on a CLN sample after GCIB sputtering. The C$1s$ line is almost completely removed, showing removal of surface lying organics and adventitious carbon. O$1s$ line narrows as all oxygen atoms now contribute to metal oxide bonding. The line shape of the Nb$3d$, Nb$4s$, and Li$1s$ peaks are unchanged. 

Finally, we use the in-situ GCIB to argon ion sputter the metalized SAW device on the "GCIB" sample, which we subsequently cool down for microwave measurements. Just as in the CLN device, we see a removal of oxygen and carbon species. This sample probes the effects of removed surface organics and adventitious carbons after metal liftoff, as well as argon ion sputtering of the surface. 

\begin{figure*}[t!]
\includegraphics[width=\linewidth]{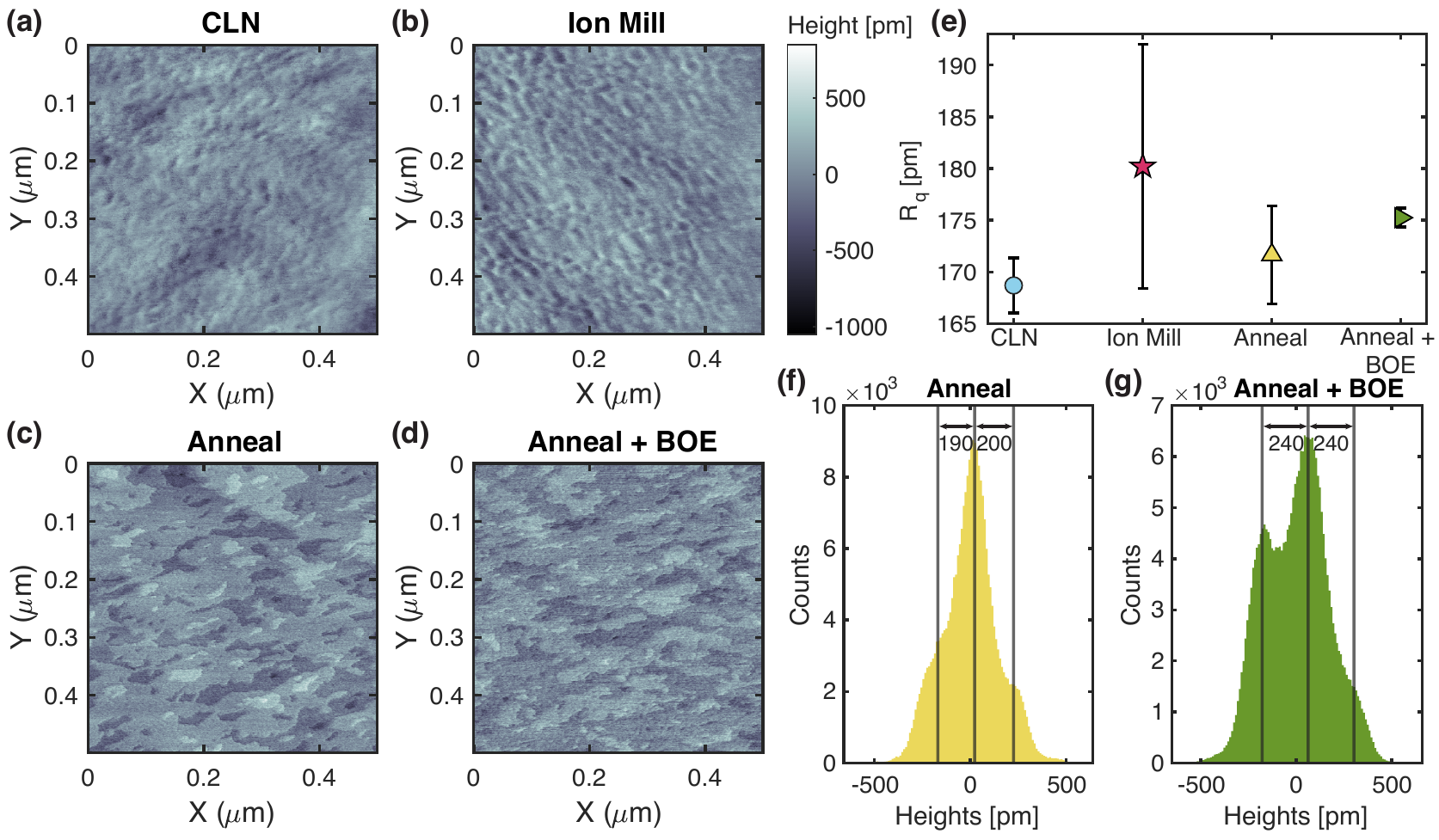}
\caption{\label{fig:fig4} \emph{AFM of various surface prepared LN substrates} (a)-(d) 0.5~$\mu$m square, 512 pixel contact mode AFM image of unprocessed congruent LN, argon ion milled LN, annealed at 500~C for 8~hr LN, and annealed and BOE dipped LN. Each image is made on x-cut material, with the X scanning direction of the AFM along crystal Z+30\degree. A change in surface morphology is noticed after annealing, and maintained after BOE dip. (e) The RMS roughness of each material process type; error bars are determined from several AFM images. (f)-(g) The histograms of the flattened anneal and anneal + BOE samples. The histograms are fit to a sum of three Gaussian peaks to determine the step heights between atomic terraces. Mean step height of the annealed sample is $200 \pm 80 \text{pm}$, and $240 \pm 80 \text{pm}$ for annealed + BOE. Uncertainty is set by the fit standard deviation of the Gaussian peaks.}
\end{figure*}

\section{Atomic Force Microscopy}
To further correlate our TLS findings with surface properties, we image several surface treated LN samples with atomic force microsocopy (AFM). We use a Bruker MultiMode-8 HR atomic force micrsocope in contact mode to image x-cut LN samples with the same surface preparation steps as the SAWs. All AFM images are taken such that the AFM scanning direction X is along crystal Z+30\degree of the LN. From the AFM scans we investigate how surface processing changes surface morphology and roughness between the CLN, Annealed, BOE and argon ion milled samples. We remove sample tilt in each AFM scan with a 1-d polynomial background subtraction prior to image processing and analysis.

Surface morphologies of the four AFM studied substrates are shown in Figure~\ref{fig:fig4} (a)-(d).  The surface morphologies of the LN samples before any treatment and after argon ion milling are similar. However, significant changes in the surface morphology are observed after the annealing process. For the annealed samples, there are groupings of higher and lower heights in isolated islands. This is likely a signature of LN surface termination relaxation from non-crystalline LN to single atomically-flat terraces of crystalline LN. We verify that there are unique step heights by further flattening these images using a three-point level on one individual atomic terrace and plot its histogram. The histograms are fit to a sum of three Gaussian peaks, where the centers of each Gaussian peak are the mean heights of each atomic layer. The mean step height of the annealed sample is $200 \pm 80~\text{pm}$, and $240 \pm 80~\text{pm}$ for annealed + BOE. Previous work using AFM to study annealed x-cut LN finds the step heights between surface terraces to be $0.24 \pm 0.2~\text{nm}$ \cite{sanna2014unraveling}, in close agreement with our findings. 

We also calculate the RMS roughness of each sample after initial 1-d polynomial background tilt removal. Argon ion milling roughened the surface most significantly, and also increased the variability of $R_q$ across the chip, seen in Figure \ref{fig:fig4} (e) as larger error bars for argon ion milled devices. The addition of annealing and the BOE dip after annealing does not change the surface roughness significantly, when compared to the control sample's RMS roughness and error bars.

\section{Results and Future Experiments}

\begin{table*}[ht]
\caption{\emph{Summary of the temperature sweep, XPS, and AFM results for each surface treatment.} Reported $F \delta^0_\text{TLS}$ is calculated from the average of all fitted resonances of each surface type. XPS numbers reported are the averaged oxygen and carbon atomic percentage ratios before performing the in-situ GCIB sputter. AFM results are reported as both the summary of surface morphology types and roughness. MgO and piranha devices are not measured in AFM. MgO and GCIB devices are not compared to other samples after XPS measurement because of other atomic species present (Mg and Al, respectively).}
\centering
\begin{ruledtabular}
\begin{tabular}{l l l l l l}
\textbf{Surface Treatment} & \textbf{$\overline{F \delta^0_\text{TLS}}$} & \textbf{Pre-GCIB O/Nb}  & \textbf{Pre-GCIB C/Nb}  & \textbf{AFM Surface Morphology} & \textbf{AFM Roughness $R_q$} [pm]\\
CLN & $5.8\times 10^{-6}$ & $3.78$ & $3.83$ & Pitted & $168.7$ \\
Annealed & $2.48\times 10^{-5}$ & $3.19$ & $0.78$ & Atomically-flat islands & $171.7$ \\
BOE & $7.7\times 10^{-6}$ & $2.91$ & $1.38$ & Atomically-flat islands & $175.3$ \\
MgO & $1.24\times 10^{-5}$ & N/A & N/A & N/A & N/A \\
Piranha & $1.06\times 10^{-5}$ & $2.97$ & $0.14$ & N/A & N/A\\
GCIB & $7.53\times 10^{-5}$ & N/A & N/A & Pitted & $180.2$\\
\end{tabular}
\end{ruledtabular}
\end{table*}

In this investigation, we explore surface treatment methods and the impact of doping on LN to understand their effects on quantum acoustic applications. We find that both thermal annealing and GCIB processing can significantly improve aspects of the surface quality as observed by two methods: AFM and XPS. Thermal annealing results in less disordered surfaces as per AFM results, whereas GCIB processed surfaces show improved characteristics, i.e., reduced carbon and associated oxygen bonding, under XPS examination.

Contrary to our expectations, however, despite improving surface quality in certain aspects, these treatments increase two-level system (TLS) density for the samples we tested. High TLS density is detrimental to the performance of quantum systems, and significant efforts have started to illuminate the effects of fabrication processes on TLS emerging in superconducting qubits~ \cite{nersisyan2019manufacturing, dunsworth2017characterization, mergenthaler2021effects}. For the case of lithium niobate mechanical resonators, our results seem somewhat contradictory as the surface quality improvement apparent in XPS/AFM is accompanied with increased TLS density for the GCIB and annealed samples. Furthermore, the tested BOE and piranha samples demonstrate that the surface quality improvements found in annealed and GCIB samples are not the cause of TLS loss increase as BOE and piranha have similar surface quality improvements (less disordered surfaces and adsorbed carbon and oxygen reduction, respectively) and a measured TLS density that is lower and closer to the control. Rather, the processes of argon ion milling and annealing seem to change the material in ways that XPS and AFM are not able to detect. This signifies a gap in our knowledge that must be addressed to optimize these systems' performance. 

Our results with MgO co-doped lithium niobate add to this puzzle. Congruently grown lithium niobate is nonstoichiometric as it forms from a melt consisting of only 48.6\% $\text{Li}_2\text{O}$ \cite{byer1970growth}; MgO co-doping fills vacancies of lithium and is an important technique used to improve LN's optical photorefractive properties. We do not observe a reduction in TLS as was suggested by earlier experiments on phononic crystals~\cite{wollack2021loss}. Our findings show a slight \emph{increase} in TLS density with MgO co-doping. 

Despite the surprising questions raised by our results, they shed light on the crucial role surfaces play in the performance of quantum acoustic resonators. We can now assert with high confidence that the resonators investigated here are surface-limited. Moreover, the results of the GCIB experiment show that though the metal-LN surface may play a role, a process affecting only the LN surface significantly changes the observed TLS density. The characterizations we have performed on these devices help us identify paths for deeper investigations to unlock the full potential of LN-based quantum acoustic devices.

Future work should consider incorporating a larger number of advanced characterization techniques to unravel the complex interplay between surface treatment, TLS density, and doping. Transmission electron microscopy (TEM) provides high-resolution imaging and analysis that can help us understand changes in LN's crystal structure near the surface due to different treatments. MgO co-doping, which did not yield the expected reduction in TLS in our experiments, could be re-evaluated with material from another vendor, akin to that from Ref.~\cite{wollack2021loss}. Furthermore, current fabrication methods using resist-based liftoff processes could be replaced by resist-free processes, potentially reducing the TLS density~\cite{tsioutsios2020free}. Alternatively, using a metal other than aluminum that can undergo more aggressive cleaning processes might reduce surface TLS.

Progress in quantum acoustics needs a more granular understanding of the effects of surface treatment techniques and material alterations on dissipation and dephasing. While unearthing unexpected results, our findings underscore the critical importance of surface properties in LN-based quantum acoustic resonators and point to a path forward for understanding and hopefully mitigating TLS to enable powerful quantum acoustic technologies.

\begin{acknowledgments}
The authors would like to thank Dr. Agnetta Cleland, Dr. Juliet Jamtgaard, Takuma Makihara, Kevin K. S. Multani, Dr. Carsten Langrock, and Professor Martin Fejer for experimental support and helpful discussions. We acknowledge funding from Amazon Web Services Inc., the David and Lucille Packard Fellowship and the Stanford University Terman Fellowship, as well as the U.S. government through the National Science Foundation CAREER award No. ECCS-1941826, the Office of Naval Research (ONR) under grant No. N00014-20-1-2422, the U.S. Air Force Office of Scientific Research (MURI Grant No. FA9550- 17-1-0002), and the U.S. Department of Energy through Grant No. DE-AC02-76SF00515 (through SLAC). E.A.W. was supported by the Department of Defense through the National Defense \& Engineering Graduate Fellowship. Part of this work was performed at the Stanford Nano Shared Facilities (SNSF) and at the Stanford Nanofabrication Facility (SNF), supported by the National Science Foundation under award ECCS-2026822. The authors wish to thank NTT Research for their financial and technical support.
\end{acknowledgments}

\appendix

\section{Reducing Diffraction Loss} \label{walkoff}

Due to the anisotropy of lithium niobate, SAWs fabricated on x-cut LN can have a non-zero beamsteering angle. To reduce diffraction losses in our SAW devices and therefore improve $Q_{res}$, we find a SAW drive direction on the LN crystal which minimizes the beamsteering angle through simulation. Using an FEM solver \cite{COMSOL}, we model an IDT unit cell on x-cut Lithium Niobate, sweeping the IDT drive direction from crystal $\text{Z} - 90\degree$ to $\text{Z} + 90\degree$. The beam steering angle $\eta$ for each crystal drive orientation is solved by calculating: 
$$ \eta = \arctan{\frac{\varoiint_{s} P_{\perp} ds / A_{\perp}}{\varoiint_{s} P_{\parallel} ds/ A_{\parallel}}},$$ where $P_{\parallel}$ and $P_{\perp}$ are the power through the parallel and perpendicular faces to the SAW drive direction, respectively, and similarly $A_{\perp}$ and $A_{\parallel}$ are the areas of each designated unit cell face. We find two drive directions on an x-cut crystal that correspond to a beamsteering angle of $0\degree$: $\text{Z} - 30\degree$ and $\text{Z} + 75\degree$, the first being the crystal orientation used in this study. 

\begin{figure}[ht!]
\includegraphics[width=\linewidth]{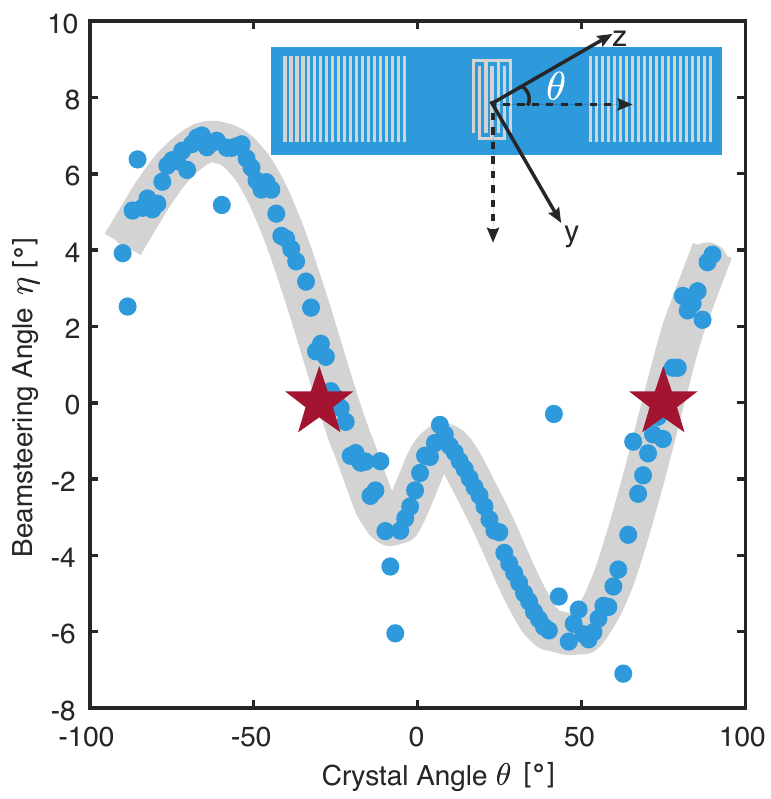}
\caption{\label{fig:figsi1} A plot of x-cut Lithium Niobate beam steering angle as a function of SAW drive crystal direction, blue is solved from FEM model, and grey is the smoothed data. The inset shows a schematic for how $\theta$ is defined, where the dotted lines represent the SAW drive direction and the solid arrows show true crystal Z and Y. The red stars represent the two identified crystal orientations with zero beam steering angle, optimum for reducing diffraction loss on x-cut LN.}
\end{figure}

\section{Dark Modes} \label{darkmodes}

\begin{figure}[ht!]
\includegraphics[width=\linewidth]{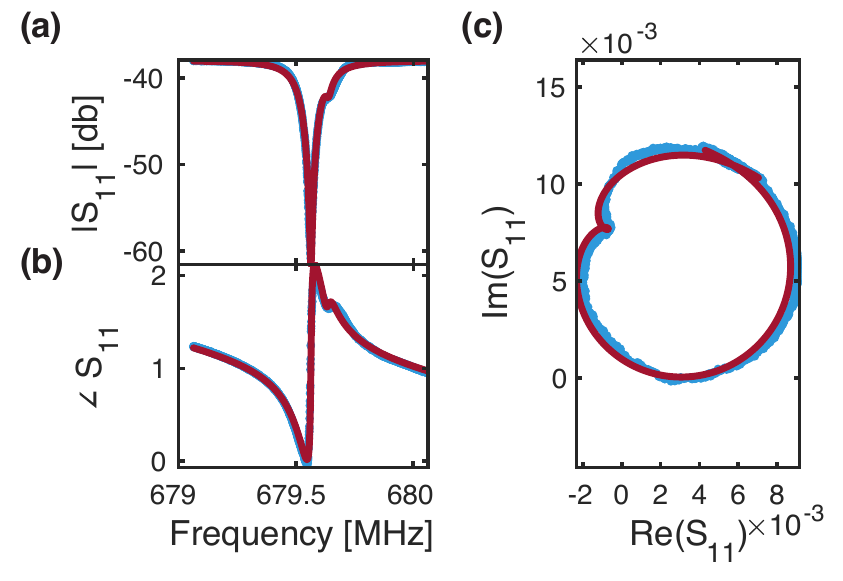}
\caption{\label{fig:figsi2} \emph{SAW Dark Mode Fit} (a) Magnitude, (b) phase and (c) imaginary vs real parts of measured (blue) and fit (red) $S_{11}$. This SAW resonance demonstrates a small dark mode 75kHz above the primary resonance ($\omega_\text{dark} = 2 \pi \times$ 679.639 MHz). This mode came from the BOE device. The data is fit to equation \ref{eq:three}. }
\end{figure}

Many of the SAW modes do not have a perfect Lorentzian line shape. We often see another more weakly coupled modes at nearby frequencies as shown in Fig
\ref{fig:figsi2}. We associate this effect with coupling of the main mode of an interest with another ``dark'' mode that is electromechanically coupled through the mode of interest. We fit this these modes more accurately to a model where a dark mode is weakly coupled to the drive through the primary SAW mode. Modes with dark modes are fit to the function
\begin{equation}
S_{11} = 1 - \frac{\kappa_e }{i \Delta+\kappa/2 + \frac{g^2} {i \Delta_b+\gamma/2}},
\label{eq:three}
\end{equation}
where $\Delta$ and $\Delta_b$ are the detunings from the drive frequency to the primary mode and dark mode, respectively, $\kappa_e$ is the external coupling rate to the primary mode, $\kappa$ is the total loss rate of the primary mode, $\gamma$ is the internal loss of the dark mode, and $g$ is the coupling rate between the primary and the dark mode. 

We use this model to fit all SAW modes in our temperature sweep measurements, to more accurately extract the resonance frequency and therefore the TLS loss product.

\section{Power Sweeps} \label{powersweeps}
We measure the devices at base temperature at a range of powers and extract intrinsic quality factors at average phonon levels ranging from $1$ to $1\times10^{10}$. Larger drive powers should saturate the TLS loss channels and therefore increase the total internal quality factor $Q_{i,\text{tot}}$, as described by the function \cite{pappas2011two}
\begin{equation}
\frac{1}{Q_{i,\text{tot}}} = \frac{F \delta^0_\text{TLS} \tanh(\frac{\hbar \omega_r}{2 k_B T})}{\sqrt{1 + \frac{\langle n \rangle}{n_c}^\beta}} + \frac{1}{Q_{i,\text{res}}}
\label{eq:two}
\end{equation}
While some modes could be fit to the $Q_{i,\text{tot}}$ and extract similar $F \delta^0_\text{TLS}$ loss to the temperature shift model, these measurements were less sensitive to TLS loss as $Q_{i,\text{res}}$ dominated the loss. $Q_{i,\text{TLS}}$ changes due to increasing drive power varied much less than the $Q_{i,\text{res}}$ changes from device to device due to metalization variation, i.e. metal finger width changes from slight differences in resist development. 

\begin{figure}[ht!]
\includegraphics[width=\linewidth]{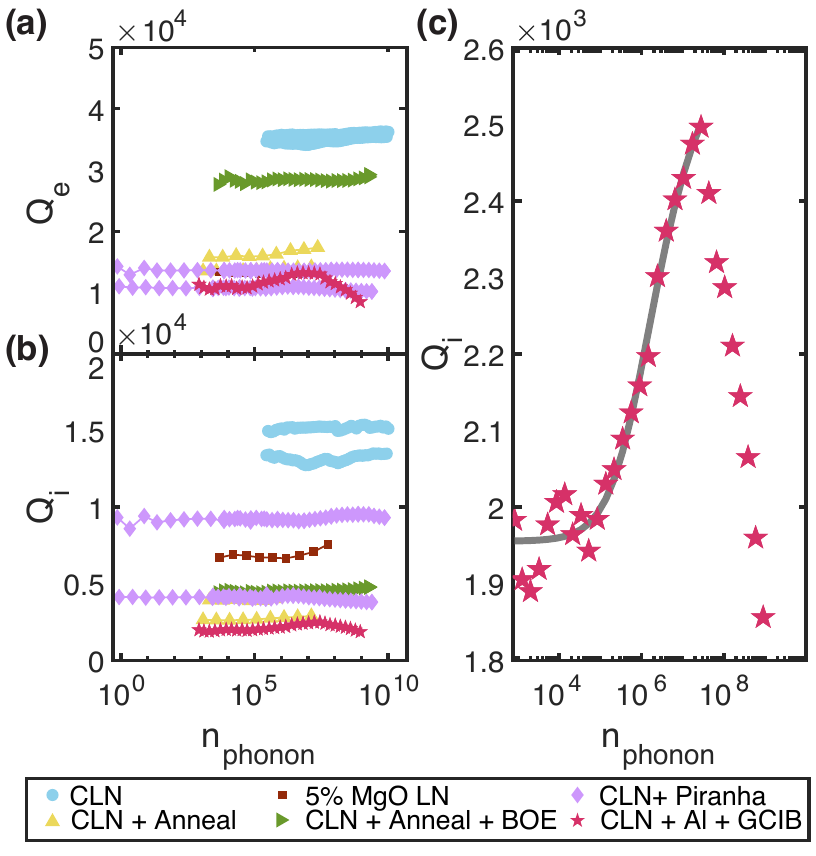}
\caption{\label{fig:figsi3} \emph{SAW Power Sweeps} (a) External and (b) internal quality factors are plotted as a function of mean phonon number for each of the surface processed SAW devices. A zoomed in image of the internal quality factor for the GCIB device is shown in figure (c). The internal quality factors are fit to  \ref{eq:two} up to an average phonon number of $2.8\times10^7$, plotted in gray. After that, the $Q_i$ decreases once again, likely due to heating or other nonlinearities.}
\end{figure}

Power sweeps of all surface treated devices are shown in Fig \ref{fig:figsi3}. The overall change in $Q_i$ from low power to high power ranged from 3.2\% in a CLN device to 30.1\% in the GCIB device. These $Q_i$ increases can be fit to equation \ref{eq:two} as seen in \ref{fig:figsi3} (c). The fit $F \delta^0_\text{TLS}$ for this device with power sweeps is $5.66\times10^{-4}$, 7.6 times larger than the $F \delta^0_\text{TLS}$ fit from the same mode via resonant frequency red shift. The difference in fit $F \delta^0_\text{TLS}$ is likely due to the fact that fit $Q_\text{res} = 2.6\times10^3$ is an order of magnitude less than $1/F \delta^0_\text{TLS} = 1.3\times10^4$, making the overall $Q_{i,\text{tot}}$ fairly insensitive to TLS power saturation. Still, where possible, the fit TLS loss from $Q_i$ follow the same general trends in surface preparation: highest TLS losses are found in GCIB devices, followed by annealed. All other surface preparations have similar TLS losses.

\bibliography{bib}

\end{document}